\documentclass[runningheads]{llncs}
\usepackage[T1]{fontenc}

\usepackage{graphicx}
\usepackage{hyperref}

\newcommand{\blindfootnote}[1]{%
  \let\thefootnote\relax%
  \footnotetext{#1}%
}

\begin{document}
\title{Prompt-Guided Patch UNet-VAE with Adversarial Supervision for Adrenal Gland Segmentation in CT Medical Images}
\titlerunning{Prompt-Guided Network for Adrenal Gland Segmentation in Medical Images}
%
\author{Hania Ghouse$^1$\orcidID{0009-0009-6619-2141}, Muzammil Behzad$^{1,2,*}$\orcidID{0000-0003-3693-4596}}
\authorrunning{H. Ghouse and M. Behzad}

\institute{$^1$King Fahd University of Petroleum and Minerals, Saudi Arabia \\ $^2$SDAIA-KFUPM Joint Research Center for Artificial Intelligence, Saudi Arabia}

\maketitle 

\begin{abstract}
Segmentation of small and irregularly shaped abdominal organs, such as the adrenal glands in CT imaging, remains a persistent challenge due to severe class imbalance, poor spatial context, and limited annotated data. In this work, we propose a unified framework that combines variational reconstruction, supervised segmentation, and adversarial patch-based feedback to address these limitations in a principled and scalable manner. Our architecture is built upon a VAE-UNet backbone that jointly reconstructs input patches and generates voxel-level segmentation masks, allowing the model to learn disentangled representations of anatomical structure and appearance. We introduce a patch-based training pipeline that selectively injects synthetic patches generated from the learned latent space, and systematically study the effects of varying synthetic-to-real patch ratios during training. To further enhance output fidelity, the framework incorporates perceptual reconstruction loss using VGG features, as well as a PatchGAN-style discriminator for adversarial supervision over spatial realism. Comprehensive experiments on the BTCV dataset demonstrate that our approach improves segmentation accuracy, particularly in boundary-sensitive regions, while maintaining strong reconstruction quality. Our findings highlight the effectiveness of hybrid generative–discriminative training regimes for small-organ segmentation and provide new insights into balancing realism, diversity, and anatomical consistency in data-scarce scenarios.

\blindfootnote{This work was supported by KFUPM grant number EC241013. The authors would also like to acknowledge the SDAIA-KFUPM Joint Research Center for Artificial Intelligence for the computational resources. \\
$^*$ indicates corresponding author. Email: 
\url{muzammil.behzad@kfupm.edu.sa}}

\end{abstract}
\section{Introduction}
Automated segmentation of small abdominal organs remains a critical yet underexplored area in medical imaging \cite{ghouse2025mosaic}-\cite{0}. While modern deep learning architectures, particularly encoder–decoder CNNs, excel at segmenting large, well-defined structures, performance deteriorates sharply when targeting small and irregularly shaped organs such as the adrenal glands \cite{1}. These organs appear with low contrast and considerable anatomical variability in CT scans, making manual annotation labor-intensive and automated segmentation prone to errors \cite{2}. Recent advances in generative modeling have introduced new opportunities to augment scarce datasets. Variational Autoencoders (VAEs) and Generative Adversarial Networks (GANs) \cite{4} have been proposed to generate realistic anatomical variations\cite{3}. Furthermore, patch-based training strategies enable networks to learn from local features, improving focus on small-region context \cite{5}.  Despite these developments, the joint use of VAE-based reconstruction, segmentation supervision, and adversarial patch-based feedback remains under-examined \cite{7}. It is unclear how injecting synthetic patches affects segmentation accuracy, reconstruction quality, and anatomical consistency across a spectrum of real-to-synthetic ratio regimes\cite{6}.  To address this gap, we propose a patch-based UNet–VAE architecture with dual-head output (reconstruction and segmentation), reinforced by a PatchGAN discriminator.
 
\section{Proposed System}
\begin{figure}[t!]
  \centering
  \includegraphics[width=1.0\textwidth]{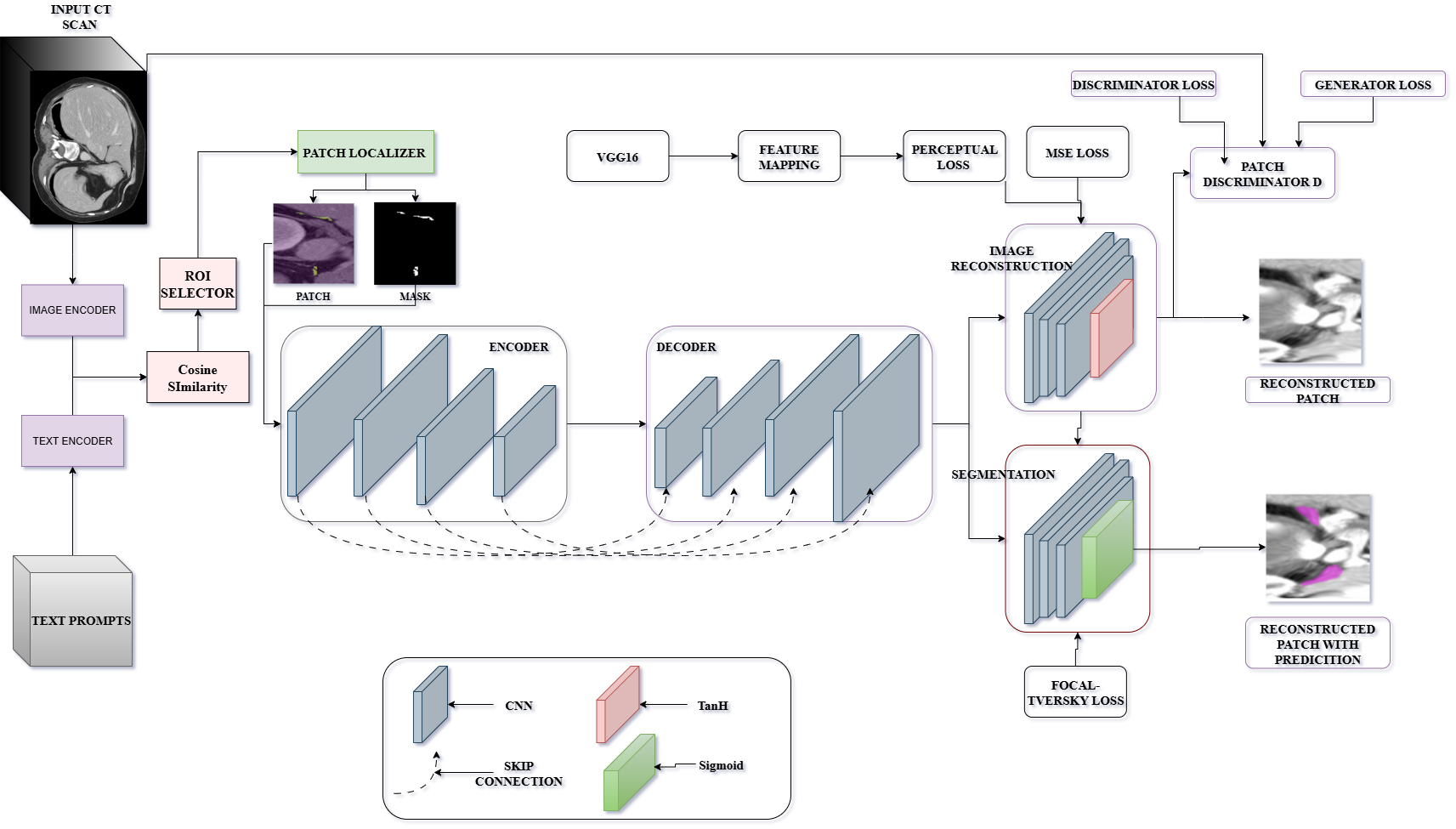}
  \caption{Architecture of the proposed UNet-VAE model for joint reconstruction and segmentation.}
  \label{fig:static-gif}
\end{figure}

We propose a patch-based UNet-VAE framework with dual-output heads and adversarial supervision, optimized for small-organ segmentation from abdominal CT images. As shown in Figure \ref{fig:static-gif}, our system begins with a text-prompt-driven region localization process. A VLM-based text encoder processes anatomical prompts, while a parallel VLM image encoder extracts visual features from the input CT scan. The cosine similarity between text and image embeddings is computed to guide an ROI selector that identifies semantically relevant spatial locations. These are passed to a Patch Localizer, which extracts localized image–mask pairs to be used during training. Each extracted patch is passed through a UNet-style encoder–decoder network that includes skip connections to preserve spatial detail. The decoder bifurcates into two parallel heads: a reconstruction head with Tanh activation that outputs a visually realistic patch reconstruction, and a segmentation head with Sigmoid activation that produces a binary mask corresponding to the organ of interest. These dual outputs are supervised using both pixel-wise and perceptual objectives. The reconstruction head is trained using a combination of pixel-level mean squared error (MSE) loss and a perceptual loss derived from frozen VGG16 feature embeddings. This enforces both low-level fidelity and high-level structural realism. To further improve anatomical plausibility, the reconstructed patch is also passed to a PatchGAN-style discriminator trained to distinguish real from synthetic patches. This introduces two additional losses: a discriminator loss that updates the PatchGAN weights and a generator loss that improves the reconstruction head’s realism via adversarial gradients. The segmentation head is supervised using the Focal Tversky Loss, which is well-suited to imbalanced segmentation problems involving small structures. During training, both heads are optimized jointly under a weighted multi-loss regime. At inference time, the model outputs a reconstructed patch and its predicted segmentation. Reconstruction quality is assessed using MSE, SSIM, and PSNR, while segmentation performance is measured using Dice, IoU, precision, recall, and Hausdorff Distance.
\section{Experimental Results and Discussions}
\begin{figure}[t]
  \centering
  \includegraphics[width=1.0\textwidth]{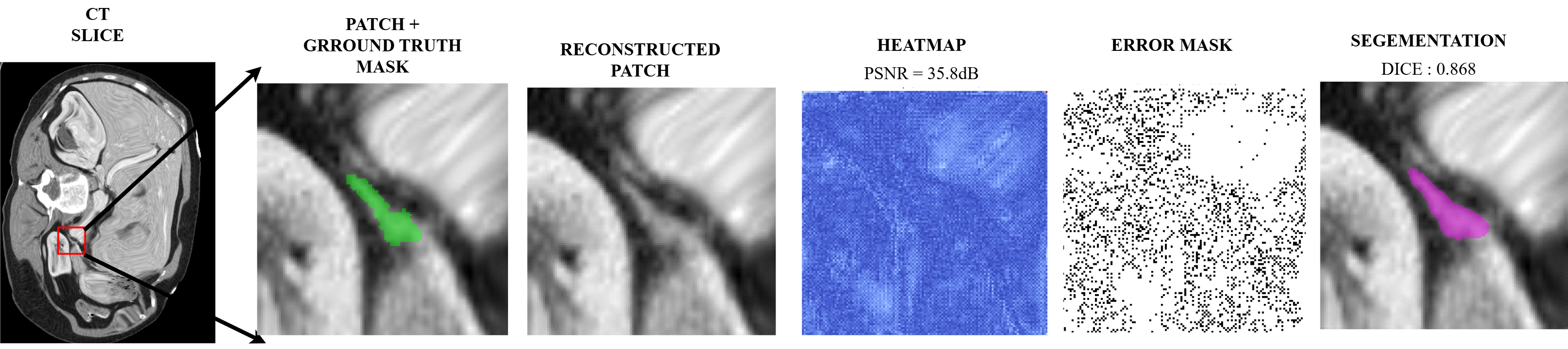}
  \caption{Visual breakdown of VAE-based reconstruction and segmentation performance on a high performing test adrenal patch.}
  \label{fig:one}
\end{figure}

\begin{figure}[t]
  \centering
  \includegraphics[width=1.0\textwidth]{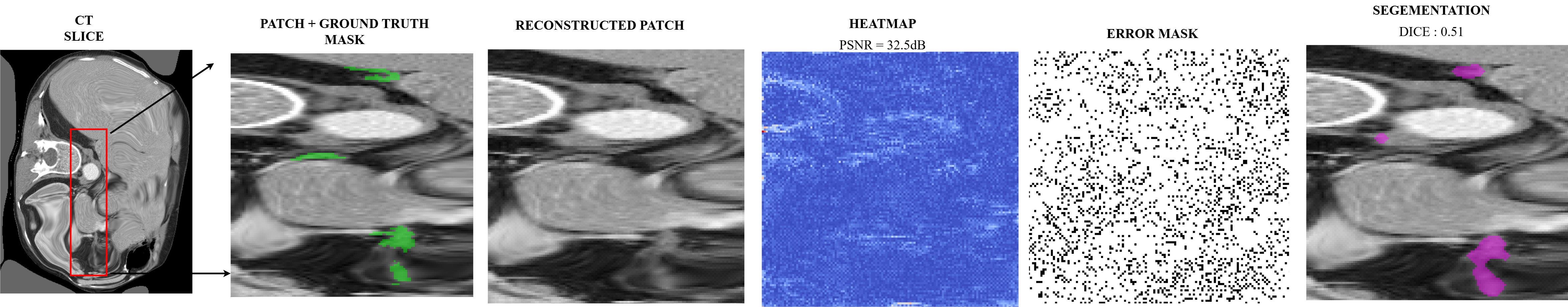}
  \caption{Visual breakdown of VAE-based reconstruction and segmentation performance on a challenging adrenal patch.}
  \label{fig:two}
\end{figure}

The model is trained using prompt-localized patches and jointly optimized for image reconstruction and mask prediction. The quantitative results are summarized in Table \ref{tab:rec_metrics} and Table \ref{tab:seg_metrics}, while the qualitative results are shown in Figure~\ref{fig:one} and Figure \ref{fig:two}. Our model achieves strong reconstruction performance, with a peak PSNR of 34.16 dB and SSIM of 0.9581, indicating high structural fidelity. The segmentation head yields a Dice coefficient of 0.884 and IoU of 0.802, demonstrating accurate organ boundary alignment despite the small size and low contrast of adrenal regions. The low Hausdorff Distance further validates the precision of predicted contours. Figure \ref{fig:one} shows a successful case where both reconstruction and segmentation are visually accurate, with minimal error and a smooth heatmap. In contrast, Figure \ref{fig:two} highlights a challenging case that involves multiple adjacent organs; Although the reconstruction is plausible, the segmentation performance drops (Dice = 0.51), showing model limitations in cluttered contexts. Across the synthetic:real ratio study, we observe that a 0.75 synthetic ratio consistently leads to the best reconstruction and segmentation trade-off. However, the model remains robust across all configurations, confirming its generalization ability and strong performance in localized adrenal segmentation. While the current study is focused on adrenal gland localization within the BTCV dataset, the proposed framework is modular and generalizable to other anatomically small structures. The influence of prompt-based selection on downstream segmentation performance also offers promising avenues for further exploration, particularly through targeted ablation studies and multi-organ evaluation.
\begin{table}[b]
  \centering
  \caption{Reconstruction metrics for different synthetic:real patch ratios.}
  \label{tab:rec_metrics}
  \begin{tabular}{c c c c c c}
    \hline
    \textbf{Ratio} & \textbf{MSE} & \textbf{MAE} & \textbf{RMSE} & \textbf{PSNR (dB)} & \textbf{SSIM} \\
    \hline
    0    & 0.002718 & 0.03980 & 0.05181 & 31.79 & 0.9314 \\
    0.25 & 0.002891 & 0.04175 & 0.05344 & 31.52 & 0.9202 \\
    0.5  & 0.002352 & 0.03687 & 0.04824 & 32.40 & 0.9194 \\
    0.75 & 0.001578 & 0.02948 & 0.03946 & 34.16 & 0.9581 \\
    1    & 0.002381 & 0.03755 & 0.04850 & 32.36 & 0.9415 \\
    \hline
  \end{tabular}
\end{table}
\begin{table}[ht]
  \centering
  \caption{Segmentation metrics for different synthetic:real patch ratios.}
  \label{tab:seg_metrics}
  \begin{tabular}{c c c c c c}
    \hline
    \textbf{Ratio} & \textbf{Dice} & \textbf{IoU} & \textbf{Precision} & \textbf{Recall} & \textbf{Hausdorff (px)} \\
    \hline
    0    & 0.8745 & 0.7887 & 0.4549 & 0.4786 & 24.73 \\
    0.25 & 0.8670 & 0.7782 & 0.4533 & 0.4760 & 25.20 \\
    0.5  & 0.8827 & 0.7996 & 0.4604 & 0.4775 & 24.26 \\
    0.75 & 0.8841 & 0.8024 & 0.4598 & 0.4790 & 24.36 \\
    1    & 0.8795 & 0.7956 & 0.4579 & 0.4782 & 24.79 \\
    \hline
  \end{tabular}
\end{table}

\section{Conclusion}
In this work, we introduce a prompt-guided, patch-based UNet-VAE framework with dual reconstruction and segmentation heads, designed specifically for robust adrenal gland segmentation in CT. By integrating anatomical text prompts,VLM-based localization, and adversarial training via PatchGAN, the model effectively learns from both real and synthetic data. Our results show consistently high Dice scores and sharp reconstructions, even for extremely small structures. Notably, the model remains stable across varying patch compositions, with peak performance at a 0.75 synthetic ratio. These findings suggest that combining reconstruction realism with segmentation supervision in a prompt-aware setting is highly effective for small-organ localization and opens the door to generalization to other anatomically complex tasks. The modular design of our framework enables seamless adaptation to broader contexts, and future investigations may further examine the interplay between prompt granularity, anatomical diversity, and segmentation fidelity.

\bibliographystyle{unsrt}
\bibliography{ref}

\begin{thebibliography}{1}

\bibitem{ghouse2025mosaic}
Hania Ghouse and Muzammil Behzad.
\newblock Mosaic: A multi-view 2.5 d organ slice selector with cross-attentional reasoning for anatomically-aware ct localization in medical organ segmentation.
\newblock {\em arXiv preprint arXiv:2505.10672}, 2025.

\bibitem{0}
Fanxing Meng, Tuo Zhang, Yukun Pan, Xiaojing Kan, Yuwei Xia, Mengyuan Xu, Jin Cai, Fangbin Liu, and Yinghui Ge.
\newblock A deep learning algorithm for automated adrenal gland segmentation on non-contrast ct images.
\newblock {\em BMC Medical Imaging}, 25(1), May 2025.

\bibitem{1}
Vlad-Octavian Bolocan and et~al.
\newblock Deep learning for adrenal gland segmentation: Comparing accuracy and efficiency across three convolutional neural network models.
\newblock {\em Applied Sciences}, 15(10):5388, May 2025.

\bibitem{2}
Michael Fayemiwo and Michael et~al.
\newblock A novel pipeline for adrenal gland segmentation: Integration of a hybrid post-processing technique with deep learning.
\newblock {\em Journal of Imaging Informatics in Medicine}, March 2025.

\bibitem{4}
Kazuhiro Koshino and et~al.
\newblock Narrative review of generative adversarial networks in medical and molecular imaging.
\newblock {\em Annals of Translational Medicine}, 9(9):821–821, May 2021.

\bibitem{3}
Revathi~S A and B~Sathish Babu.
\newblock Synthesizing realistic knee mri images: A vae-gan approach for enhanced medical data augmentation.
\newblock {\em International Journal of Advanced Computer Science and Applications}, 15(11), 2024.

\bibitem{5}
Peter Klages and et~al.
\newblock Patch‐based generative adversarial neural network models for head and neck mr‐only planning.
\newblock {\em Medical Physics}, 47(2):626–642, December 2019.

\bibitem{7}
Jingkun Chen and et~al.
\newblock Unsupervised patch-gan with targeted patch ranking for fine-grained novelty detection in medical imaging, 2025.

\bibitem{6}
Ali~Q Saeed and et~al.
\newblock Synthesizing retinal images using end-to-end vaes-gan pipeline-based sharpening and varying layer.
\newblock {\em Multimedia Tools and Applications}, 83(1):1283–1307, October 2023.

\end{thebibliography}

\end{document}